\documentclass{ws-ijmpa}

\usepackage{url}
\usepackage[super]{cite}
\usepackage{xcolor}
\usepackage[verbose,hypertexnames=false]{hyperref}
\hypersetup{colorlinks=false,allbordercolors=blue,pdfborderstyle={/S/U/W 1}}
\begin{document}

\markboth{Zhibo Tao et al.}{Invisible decays of light mesons}

\catchline{}{}{}{}{}
%
%%%%%%%%%%%%%%%%%%%%%%%%%%%%%%%%%%%%%%%%%%%%%%%%%%%%%%%%%%%%%%%%%%%%

\title{Search for invisible decays of light mesons via \\$J/\psi \to VP$ $(V=\omega/\phi,P=\eta/\eta')$ decays at STCF
%{For the title, try not to use more than
%3 lines. Typeset the title in 10 pt roman, boldface.}
}

\author{Zhibo Tao$^*$, Yihang Xia, Vindhyawasini Prasad, \\
Xu Gao and Weimin Song
%\footnote{
%Typeset names in 8 pt roman. Use the footnote to indicate the
%present or permanent address of the author.}
}

\address{College of Physics, Jilin University\\
Changchun, 130012, China
%\footnote{
%State completely without abbreviations, the affiliation and
%mailing address, including country. Typeset in 8 pt italic.}
\\
$^*$taozb25@mails.jlu.edu.cn}

% \author{Second Author}

% \address{Group, Laboratory, Address\\
% City, State ZIP/Zone, Country\\
% second\_author@domain\_name}

\maketitle

\begin{history}
\received{Day Month Year}
\revised{Day Month Year}
\accepted{Day Month Year}
\published{Day Month Year}
\end{history}

\begin{abstract}
% The abstract should summarize the context, content
% and conclusions of the paper in less than 200 words. It should
% not contain any references or displayed equations. Typeset the
% abstract in 8 pt roman with baselineskip of 10 pt, making
% an indentation of 1.5 pica on the left and right margins.
% Based on an inclusive Monte Carlo sample of $1.3 \times 10^{9}$ $J/\psi$ events at STCF, 
% we generate projeted toy data to perform the sensitivity study of the invisible decays of 
% light mesons ($V=\omega/\phi,P=\eta/\eta'$) at expected $3.4\times10^{12}$ $J/\psi$
% evnets per year data taking level via $J/\psi \to VP$ decays. 
We present a preliminary feasibility study of searches for invisible decays of light mesons 
via $J/\psi \to VP$ $(V=\omega/\phi,P=\eta/\eta')$ using a traditional analytical method
at the proposed Super $\tau$-Charm 
facility (STCF) which is expected to accumulate $3.4\times10^{12}$ $J/\psi$ events per year, 
based on an inclusive Monte Carlo sample of $1.3 \times 10^{9}$ $J/\psi$ events.
The upper limits on the invisible 
decay branching fractions at the 90\% confidence level are set as 
$\mathcal{B}(\omega \to invisible) < 3.7 \times 10^{-7}$, 
$\mathcal{B}(\phi \to invisible) < 8.9 \times 10^{-7}$,
$\mathcal{B}(\eta \to invisible) < 1.8 \times 10^{-7}$ and
$\mathcal{B}(\eta' \to invisible) < 4.1 \times 10^{-7}$, respectively, using a projected toy data 
corresponding to the expected STCF statistics. By using the machine 
learning technique such as Deep Learning, the upper limit may be further improved to approach
theoretical predictions for light dark matter.
\end{abstract}


\ccode{PACS numbers: 03.65.$-$w, 04.62.+v}

\section{Introduction}
The invisible decays of light mesons, in which the final state particles
are not observable in the detector, may provide a unique probe to search for
new physics beyond the Standard Model (SM). Although it is possible that
quarkonium states ($q\bar{q}$) annihilate into a neutrino pair ($\nu\bar{\nu}$)
via a virtual $Z^{0}$ boson, this process is so rare that is beyond the scope of existing 
collider experiments.\cite{Chang1998OnTI}
Consequently, any observed significant excess of invisible decays would
constitute unambiguous evidence for new physics. In particular, the presence of 
light dark matter (LDM) particles $\chi$ \cite{PhysRevD.103.075005,PhysRevD.70.023514,JHEP09}
may enhance the branching fraction of the 
invisible decays by several orders of magnitude,\cite{PhysRevD.72.103508} reaching the level of current planned
experimental sensitivity, such as Super $\tau$-Charm facility (STCF) experiment.

LDM particles annihilating into electron-positron pairs might be 
one of the sources of the 511 keV $\gamma$-ray emission line from the galactic 
center observed by the INTEGRAL satellite.\cite{Knodlseder:2003sv,PhysRevLett.94.171301,PhysRevLett.96.211302} The LDM particles may satisfy the 
constraints from the INTEGRAL measurement and relic abundance to account for 
the nonbaryonic dark matter\cite{DelPopolo:2013qba} of the universe, if dark matter annihilates via a new light 
gauge boson U\cite{PhysRevD.79.015014} or  scalar dark matter annihilates through the exchange of 
heavy fermions\cite{PhysRevD.103.075005,PhysRevD.70.023514}. Assuming the same cross section for
the time reversed processes, $\sigma(q\bar{q}\to\chi\chi)\simeq\sigma(\chi\chi\to 
q\bar{q})$, the branching fraction of invisible decays of $V$ ($V=\omega,\phi$) 
and of $\eta (\eta')$ can reach $\sim 10^{-8}$ and $\sim 10^{-5}(10^{-7})$, 
respectively.\cite{mcelrath2007lighthiggsesdarkmatter}

The BESIII experiment has reported the upper limits on the branching fractions 
of $V\to invisible$ ($V=\omega,\phi$) via $J/\psi \to V\eta$ decays and 
$P\to invisible$ ($P=\eta,\eta'$) via $J/\psi \to \phi P$ decays, using 
$(1310.6 \pm 7.0) \times 10^{6}$ and $(225.3 \pm 2.8) \times 10^{6}$ $J/\psi$ 
events, respectively.\cite{BESIII:2018bec,PhysRevD.87.012009} 
The future STCF experiment is expected to collect $3.4\times 10^{12}$ $J/\psi$
events per year,\cite{Achasov:2023gey} providing a great opportunity to search for the invisible meson 
decays. In this paper, we report the expected sensitivity of these invisible 
decays by following the traditional analytical method.

\section{The STCF experiment and Monte Carlo simulation}
STCF is a proposed electron-positron collider
in China with a center-of-mass energy ranging from 2 to 7 GeV and a peak luminosity of 
$0.5 \times 10^{35} \mathrm{cm^{-2}s^{-1}}$, as an upgrade of the 
present $\tau$-Charm factory BEPCII. 
% It's expected to produce a data sample about 
% a factor of 100 larger than the BEPCII, providing a unique platform for searching 
% physics beyond the SM. 
It is composed of an inner tracker (ITK) and a main drift 
chamber (MDC) for charged particle track and momenta measurement, a DIRC-like time-of-flight (TOF) detector for particle identification (PID)
in the barrel, a detector for PID in the end-cap, an electromagnetic calorimeter (EMC) for photon and electron 
energy measurement and a muon detector (MUD) for muon identification. The design options for the detector for PID in the end-cap include RICH, ASHIPH 
and DTOF.

A Monte Carlo (MC) sample of $1.3 \times 10^{9}$ inclusive $J/\psi$ events 
is generated for the sensitivity study. The MC simulation is performed with GEANT4.\cite{AGOSTINELLI2003250}
The known $J/\psi$ decay modes are generated by the 
StcfEvtGen generator package with the branching fractions taken from the Particle 
Data Group (PDG),\cite{ParticleDataGroup:2024cfk} while the remaining unknown $J/\psi$ decay modes are generated by
the LUNDCHARM\cite{PhysRevD.62.034003} generator. 
% The production of the $J/\psi$ resonance via $e^{+}e^{-}$ 
% annihilation is simulated by the KKMC including the effects of the beam energy 
% spread and initial state radiation (ISR). 
% We use a helicity amplitude model for the
% $J/\psi \to V \eta$ and $J/\psi \to \phi P$ decays, an $\omega$ Dalitz plot 
% distribution model for the $\omega \to \pi^{+}\pi^{-}\pi^{0}$ decay, an $\eta$
% Dalitz plot distribution model for the $\eta \to \pi^{+}\pi^{-}\pi^{0}$ decay, a 
% vector meson decaying to a pair of scalar particles model for the 
% $\phi \to K^{+}K^{-}$ decay, and a phase space model for $V \to \nu \bar{\nu}$ and
% $P \to \nu \bar{\nu}$ decays.
Signal events are modeled using helicity amplitude model for $J/\psi \to VP$ decays, 
$\omega$ Dalitz plot distribution model for $\omega \to \pi^{+}\pi^{-}\pi^{0}$ decay,
$\eta$ Dalitz plot distribution model for
$\eta \to \pi^{+}\pi^{-}\pi^{0}$ decay, vector meson decaying to a pair of scalar particles model for
$\phi \to K^{+}K^{-}$ decay, and phase space model for other decay modes.

\section{Analysis Strategy}
We search for the invisible decays of light mesons ($\omega,\phi,\eta,\eta'$) via 
the two-body $J/\psi \to VP$ $(V=\omega/\phi,P=\eta/\eta')$ decays. For vector mesons V, we study the $J/\psi \to V \eta$ decay, 
in which the $\eta$ meson is reconstructed from $\pi^{+}\pi^{-}\pi^{0}$ decay mode. The more dominant decay mode
$\eta \to \gamma\gamma$ is not used due to the huge background contamination.
For pseudoscalar mesons P, we study the $J/\psi \to \phi P$ decay, in which the $\phi$ candidates can be  
reconstructed easily and cleanly from its $K^{+}K^{-}$ decay mode. 
In both cases, the mass 
distribution of the system recoiling against the tagged candidates is used to 
search for the invisible decays.

To cancel different sources of system uncertainty, the visible decays of 
$\omega \to \pi^{+}\pi^{-}\pi^{0}$, $\phi \to K^{+}K^{-}$ from $J/\psi \to V\eta$
decays and the decay modes of $\eta \to \gamma\gamma$, $\eta' \to \gamma\gamma$ 
from $J/\psi \to \phi P$ decays are constructed, respectively. The ratio of the 
branching fraction of the invisible decays to that of the visible decays of $X$ 
($X=V,P$) mesons is measured by
\begin{equation}
	\frac{\mathcal{B}(X \to \mathrm{invisible})}{\mathcal{B}(X \to \mathrm{visible})} = 
	\frac{N_{\mathrm{sig}}^{\mathrm{invisible}} \cdot \epsilon^{\mathrm{visible}}}
	{N_{\mathrm{sig}}^{\mathrm{visible}} \cdot \epsilon^{\mathrm{invisible}}}
	\label{e1}
\end{equation}
where $N_{\mathrm{sig}}^{\mathrm{invisible}}$ and $N_{\mathrm{sig}}^{\mathrm{visible}}$ are the number of signal 
events for invisible and visible decays, respectively, 
while $\epsilon^{\mathrm{invisible}}$ and $\epsilon^{\mathrm{visible}}$ are the corresponding 
detection efficiencies.

\section{Sensitivity estimation}
The Charged tracks that originate within $\pm10$ cm of the interaction point (IP)
in the beam direction and within $\pm1$ cm in the plane perpendicular to the beam
are selected. The polar angle $\theta$ of the charged tracks must satisfy 
$\left| \cos \theta \right| < 0.93$. The Global PID algorithm, which is developed 
based on machine learning using multiple variables reconstructed in each subdetector, is used in this analysis. A charged pion is identified
by requiring the probability of its pion hypothesis to be greater than the kaon 
and electron hypotheses. The same applies for the kaon.

The electromagnetic showers are reconstructed from clusters of energy deposited 
in the EMC. The shower energies are required to be greater than 25 MeV for the 
barrel region ($\left| \cos \theta \right| < 0.8325$) and 50 MeV for the end-cap
region ($0.8325 < \left| \cos \theta \right| < 0.9445$). The angle between a 
photon candidate and the nearest extrapolated track in the EMC should be greater 
than $10^{\circ}$. A $\pi^{0}$ candidate is reconstructed from a photon pair 
candidate, and the 2-photon invariant mass is constrained to the nominal value
of the $\pi^{0}$ meson\cite{ParticleDataGroup:2024cfk} by performing a kinematic fit.

The analysis is performed using $1.3 \times 10^{9}$ inclusive $J/\psi$ MC events.
To simulate the expected scenario of the STCF experiment, we generate the
projected toy data based on the analysis result.

\subsection{The invisible decays of $\omega$ and $\phi$ mesons}
We require that the invisible decay event candidate have no charged tracks besides
the two oppositely charged pions produced in the decay 
$\eta \to \pi^{+}\pi^{-}\pi^{0}$. A vertex fit is performed on these two charged 
tracks to ensure that they originate from a common vertex. The 
$\eta \to \pi^{+}\pi^{-}\pi^{0}$ decay process is reconstructed for which the 
$\pi^{+}\pi^{-}\pi^{0}$ invariant mass ($M_{\pi^{+}\pi^{-}\pi^{0}}$) is closest to 
the nominal mass of $\eta$ meson.\cite{ParticleDataGroup:2024cfk} Then the $M_{\pi^{+}\pi^{-}\pi^{0}}$ distribution
of the $\eta$ candidate is required to be within the range of [0.52, 0.57] 
$\mathrm{GeV}/c^{2}$. 
%To remove the combinatorial backgrounds from events, 
$E_{\gamma}^{\mathrm{extra}}$ is required to be less than 0.2 GeV, where 
$E_{\gamma}^{\mathrm{extra}}$ is the sum of the energies of all extra photons which are 
not used for $\pi^{0}$ reconstruction. The expected distribution of $E_{\gamma}^{\mathrm{extra}}$
for the signal is expected to peak at 0 while for backgrounds it does not.
In order to further remove the backgrounds 
from $J/\psi \to X \eta$, 
the polar angle of the system against the selected $\eta$
candidate, $\theta_{\mathrm{recoil}}$, should satisfy 
$\left|\cos\theta_{\mathrm{recoil}}\right| < 0.7$ 
, where X could be any final state emitted 
in the region that is not covered by the detector acceptance.

\begin{figure}[t]
\centerline{\includegraphics[width=6cm]{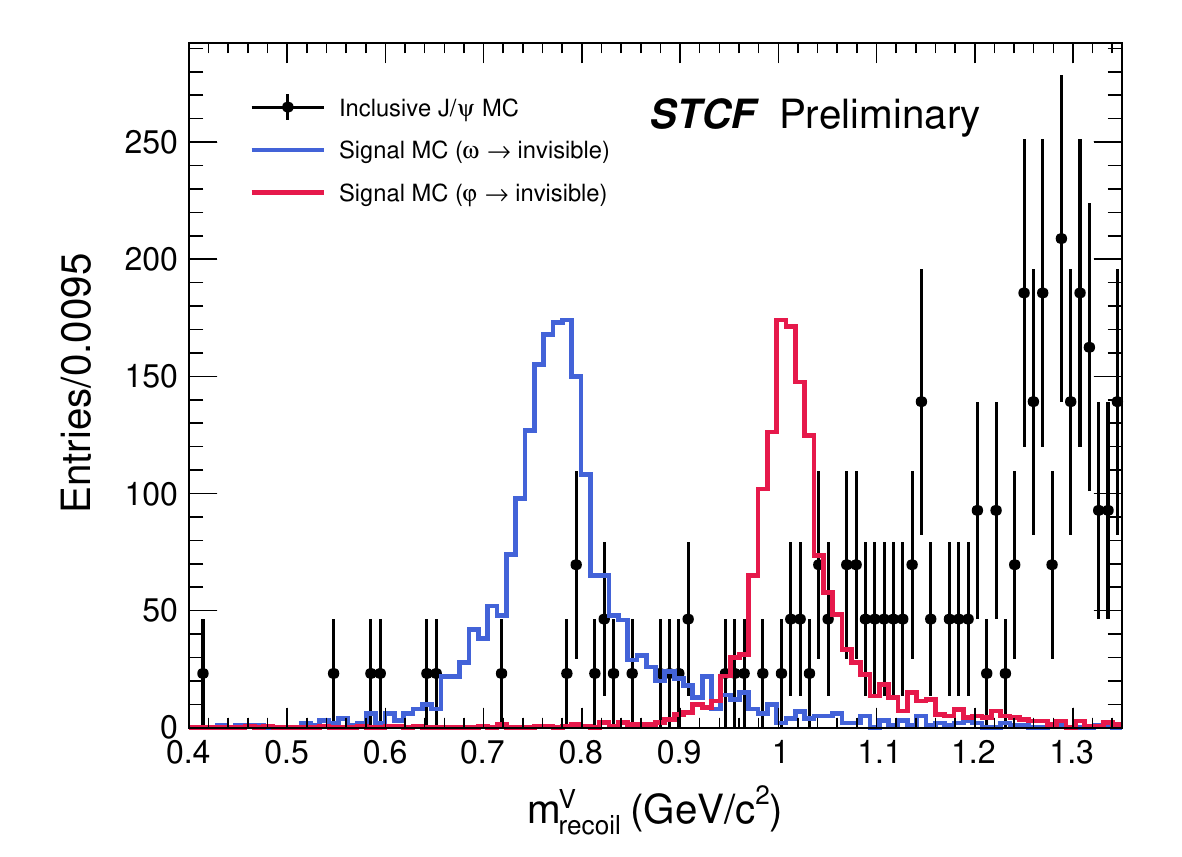}}
\caption{Invariant mass recoiling against the selected $\eta$ candidate 
($m_{\mathrm{recoil}}^{V}$) for inclusive $J/\psi$ MC samples (black dots with 
error bars) and signal MC samples (blue and red histograms for $\omega$ and
$\phi$, respectively).}
\label{f1}
\end{figure}

Maximum likelihood (ML) fits are performed on the $m_{\mathrm{recoil}}^{V}$ distribution to 
extract the signal yield of the invisible decays of $\omega$ and $\phi$ mesons, 
where $m_{\mathrm{recoil}}^{V}$ is the invariant mass of the system recoiling against the 
selected $\eta$ candidate. 
% It's defined as 
% $m_{\mathrm{recoil}}^{V} \equiv \sqrt{(E_{\mathrm{cm}} - E_{\pi^{+}\pi^{-}\pi^{0}})^2 - 
% P_{\pi^{+}\pi^{-}\pi^{0}}^2}$, where $E_{cm}$ is the CM energy, and 
% $E_{\pi^{+}\pi^{-}\pi^{0}}$ and $P_{\pi^{+}\pi^{-}\pi^{0}}$ are the energy and 
% momentum of the $\pi^{+}\pi^{-}\pi^{0}$ system in the CM frame, respectively.
The $m_{\mathrm{recoil}}^{V}$ distribution of the inclusive $J/\psi$ MC event candidates
after all selection criteria within the range of [0.40, 1.35] $\mathrm{GeV}/c^{2}$ is 
shown in Fig.~\ref{f1}. The possible peaking background comes from the decay 
$J/\psi \to V \eta$ with visible V meson decay. The numbers of the peaking 
backgrounds are evaluated to be 0.08 for $J/\psi \to \omega \eta$ and 1.14 for 
$J/\psi \to \phi \eta$ with negligible uncertainty. 

\begin{figure}[t]
\centering
\includegraphics[width=6cm]{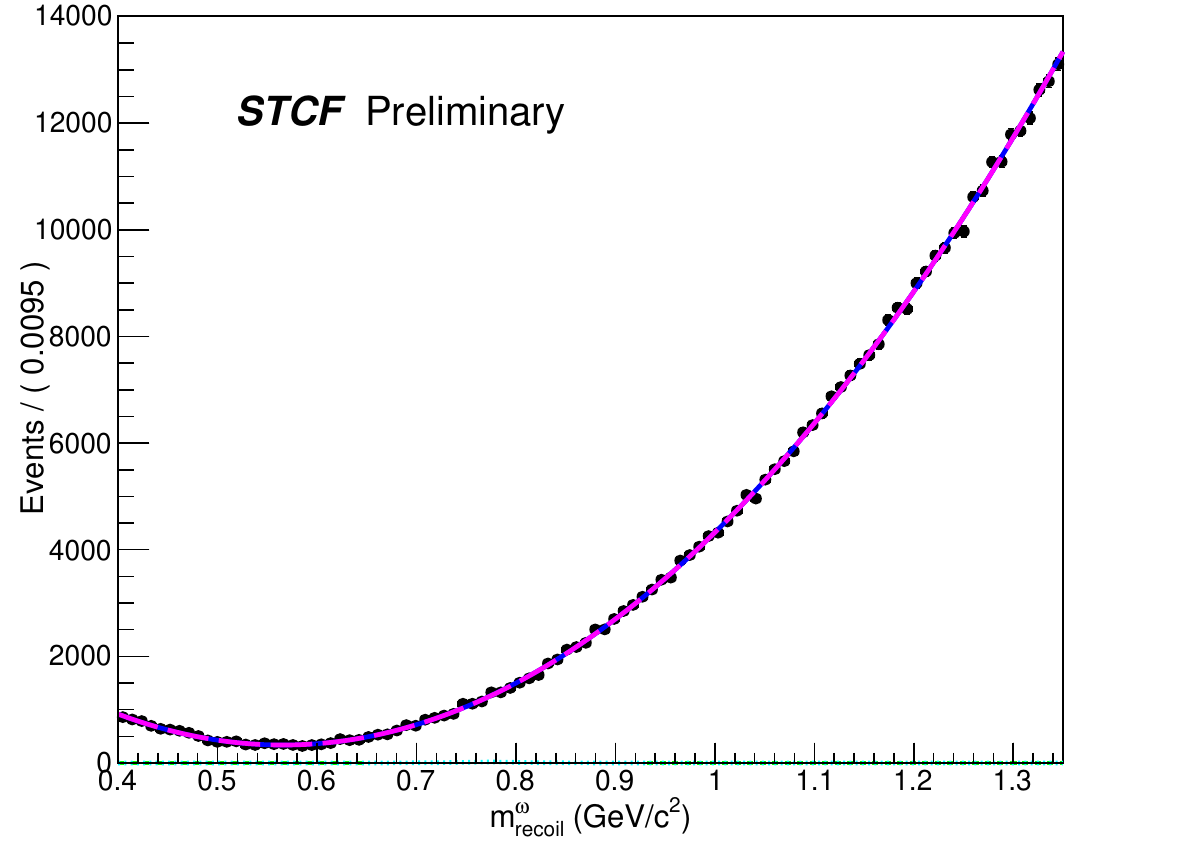}
\includegraphics[width=6cm]{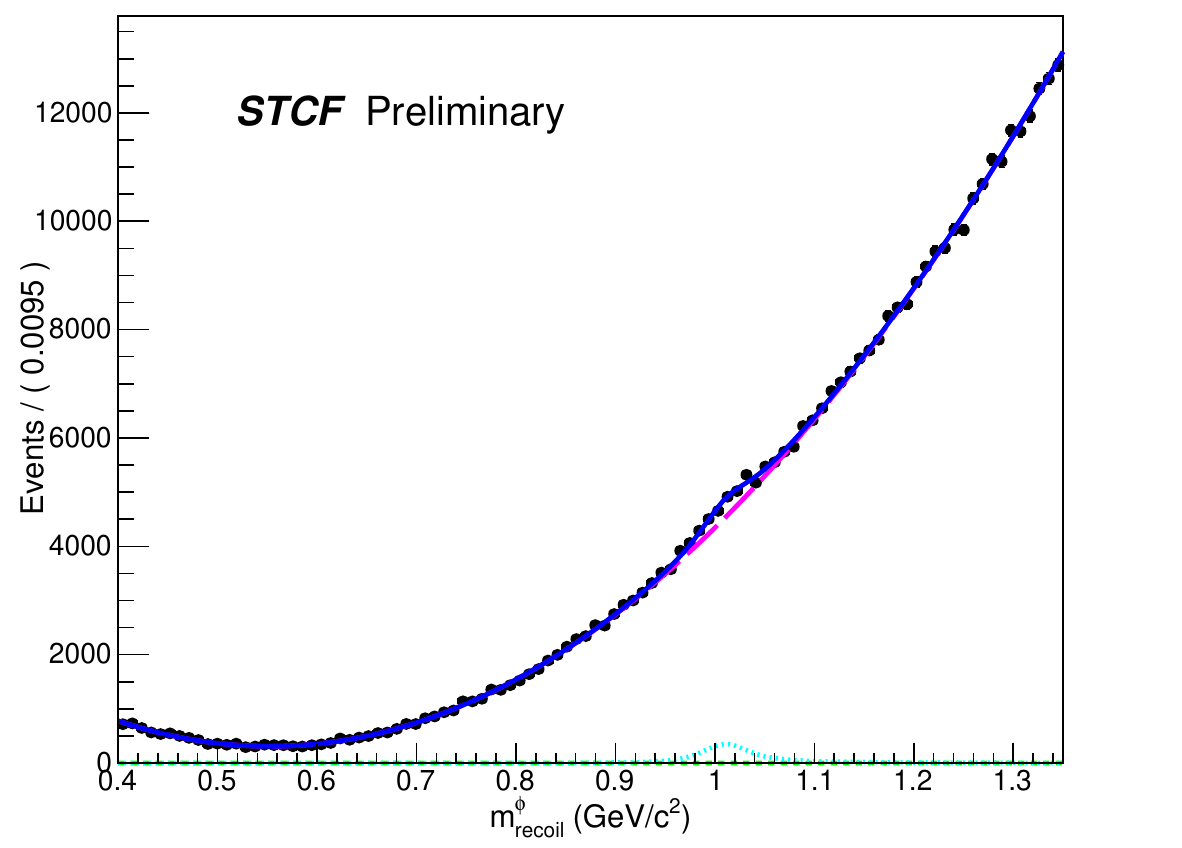}
\caption{Fit to the $m_{recoil}^{V}$ distribution for $\omega$ (left) and $\phi$
(right) signals. The projected toy data sample is shown by the dots with error bars.
The solid blue, dashed green, dotted cyan and long-dashed pink curves represent the 
overall fit results, signal, peaking and non-peaking backgrounds, respectively.}
\label{f2}
\end{figure}

The signal and peaking background probability density functions (PDFs) of the 
recoil mass distribution in the invisible decays of V meson 
is described by their MC simulated 
shapes, while the non-peaking background PDF is described by a second order
polynomial function. 
In the fit, the number of peaking background events is fixed,
while the parameters of the non-peaking background PDF and the yields for signal 
and non-peaking background events are free parameters. 
The ML fit yields are $N_{sig} 
= 0.05 \pm 4.57$ events for $\omega \to invisible$ decay and $N_{sig} = -7.40 \pm
4.53$ events for $\phi \to invisible$ decay. 
% The signal yields are consistent with 
% 0, which is reasonable since the inclusive $J/\psi$ MC samples don't include the 
% V meson invisible decays. 
% To generate 
% the projected toy data, the final values of the fit is used to set the initial values of 
% the PDF parameters and the number of background events, while the initial value of 
% signal yield is kept at 0. 
% By keeping the signal yields at 0, the total PDF is used to generate the projected toy data 
% and the ML fit yields of the projected toy data 
The total PDF obtained in this study, assuming a null signal, is used to generate the projected 
toy data, and the ML fit to the projected data yields
$N_{sig} = -199.12 \pm 238$ events for 
$\omega \to invisible$ decay and $N_{sig} = 17.56 \pm 305$ events for 
$\phi \to invisible$ decay. 
The fit results on these toy data are shown in Fig.~\ref{f2}. 
% The fitted $m_{recoil}^{V}$ of the projected toy data are also 
% shown in Fig.. 
The corresponding signal detection efficiencies
% , estimated with the 
% MC simulation, 
are 21.22\% and 20.85\% for $\omega$ and $\phi$ invisible decays, 
respectively.

\subsection{The invisible decays of $\eta$ and $\eta^{\prime}$ mesons}

The event candidates of the invisible decays of the P mesons are required to have 
exactly two oppositely charged tracks sharing a common vertex by performing a 
vertex fit. The charged tracks are assumed to be kaons, and the invariant mass of 
$K^{+}K^{-}$ is required to be in the range of [1.01, 1.03] $\mathrm{GeV}/c^{2}$. 
We also require that the angle between photons and the system recoiling against the
$\phi$ candidate $\theta_{\mathrm{recoil}\phi,\gamma}$, should satisfy 
$\theta_{\mathrm{recoil}\phi,\gamma} > 1$. Furthermore, the polar angle of the system 
recoiling against the $\phi$ candidate $\theta_{\mathrm{recoil}}$, is also required to satisfy
$\left|\cos \theta_{\mathrm{recoil}}\right| < 0.7$.

\begin{figure}[t]
\centerline{\includegraphics[width=6cm]{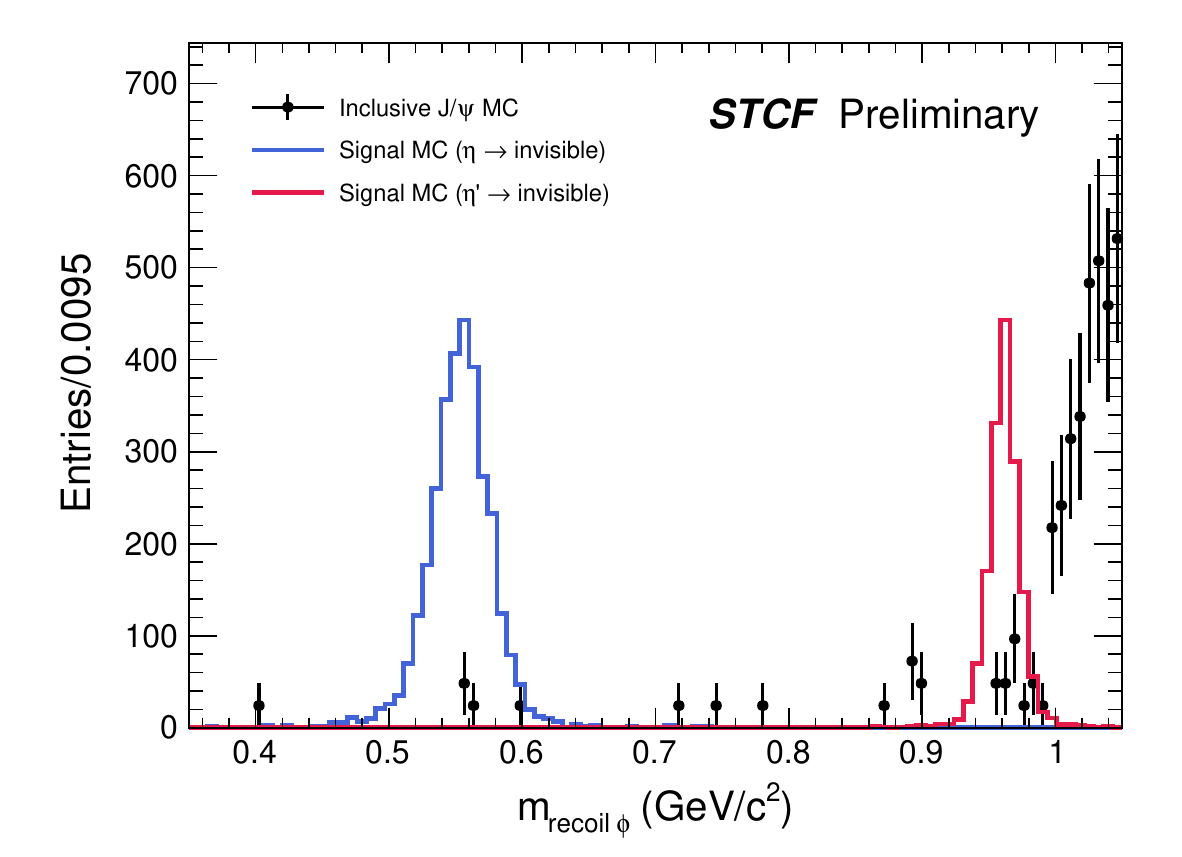}}
\caption{Invariant mass recoiling against the selected $\phi$ candidate 
($m_{\phi}^{\mathrm{recoil}}$) for inclusive $J/\psi$ MC samples (black dots with 
error bars) and signal MC samples (blue and red histograms for $\eta$ and 
$\eta^{\prime}$, respectively).}
\label{f5}
\end{figure}

After all selection criteria are applied, the corresponding signal detection 
eﬀiciencies are 31.93\% and 31.58\% for $\eta$ and 
$\eta^{\prime}$ invisible decays, respectively. The $m_{\phi}^{\mathrm{recoil}}$ 
distribution of the inclusive $J/\psi$ MC candidates with the range of [0.35, 1.05] 
$\mathrm{GeV}/c^{2}$ is shown in Fig.~\ref{f5}, where $m_{\phi}^{\mathrm{recoil}}$ is the 
invariant mass of the system recoiling against the selected $\phi$ candidate. 

In the $\eta$ signal region ($m_{\phi}^{\mathrm{recoil}} \in [0.35, 0.75] \mathrm{GeV}/c^{2}$), 
the signal PDF is described by the MC simulated shape and the background PDF is 
modeled with a first order polynomial function.
% So the invariant mass of $K^{+}K^{-}$ was was replaced with [1.04, 1.30] 
% $\mathrm{GeV}/c^{2}$ in order to obtain the sideband data. In the ML fit, the 
% shape of sideband data is modeled with a first-order Chebychev polynomial function. 
The ML fit is performed and the signal events $N_{sig} = 2.81 \pm 1.98$. 
% Using the same method, 
% The upper limit at the 90\% C.L. of generated projected toy data is $N_{up}^{\eta \rightarrow 
% invisible} = 73.59$. 

\begin{figure}[t]
\centering
\includegraphics[width=6cm]{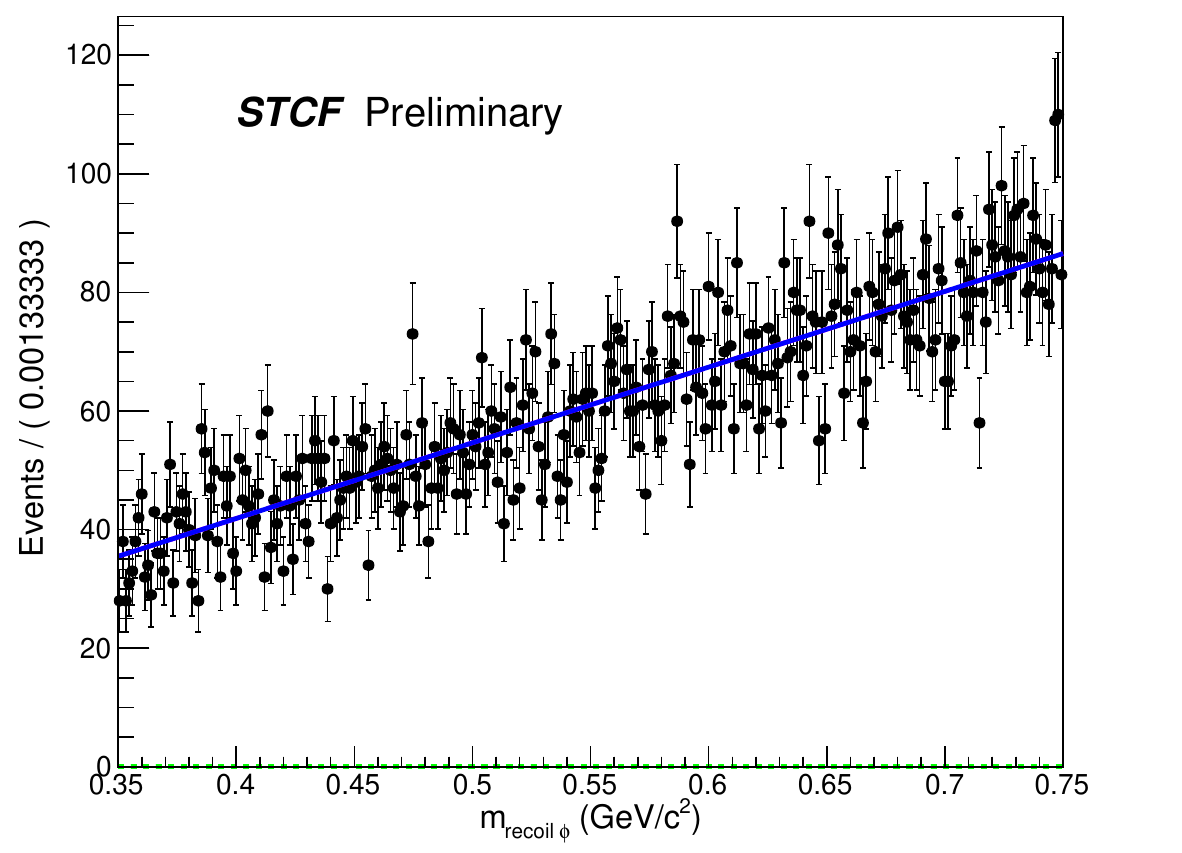}
\includegraphics[width=6cm]{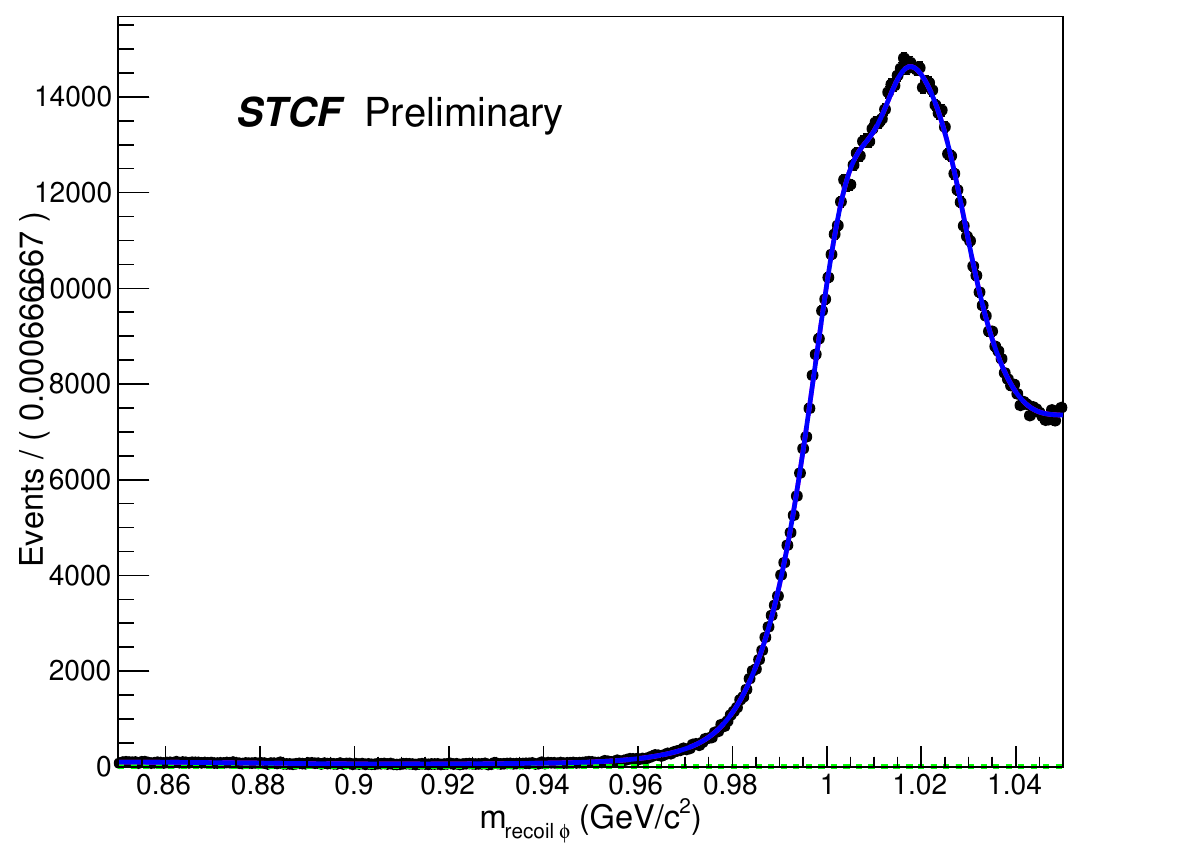}
\caption{Fit to the $m_{\phi}^{\mathrm{recoil}}$ distribution for $\eta$ (left) and 
$\eta^{\prime}$ (right) signals. The projected toy data sample is shown by the 
dots with error bars. The solid blue, long-dashed pink, and dashed green curves 
represent the overall fit results, signal and backgrounds, respectively.}
\label{f6}
\end{figure}

For the $\eta^{\prime}$ case ($m_{\phi}^{\mathrm{recoil}} \in [0.75, 1.05] \mathrm{GeV}/c^{2}$), the 
dominant backgrounds from
$J/\psi \rightarrow \phi f_{0}(980)$ ($f_{0}(980) \rightarrow K_{L}K_{L}$) and 
$J/\psi \rightarrow \phi K_{L}K_{L}$
are described by their MC simulated shapes in the ML fit and the shape of the remaining 
background is modeled with a first order polynomial function. The signal yield 
$N_{sig} = 0.00 \pm 1.35$. 

% And $N_{up}^{\eta^{\prime} \rightarrow invisible} = 88.36$ 
% is obtained. 

By following the same method of projected toy data generation as described in the previous section, 
the ML fit yields of the projected toy data are $N_{sig} = 0.0 \pm 33.2$ and $N_{sig} = 20.4 \pm 36.4$ 
for $\eta$ and $\eta'$ invisible decays, respectively. The fit results of $\eta$ and $\eta'$ invisible 
decays on toy data are shown in Fig.~\ref{f6}.

\subsection{The decay mode $J/\psi \to \omega \eta, \omega \to \pi^{+}\pi^{-}\pi^{0}, \eta \to \pi^{+}\pi^{-}\pi^{0}$}
For this decay mode, we require four charged pions with net charge zero and at 
least two independent $\pi^{0}$ candidates. The four charged tracks are required to
originate from a common vertex by performing a vertex fit. The total energy 
($E^{tot}$) of the selected candidates is required to satisfy $E^{tot} > 2.95$ GeV.
For a selected 2$(\pi^{+}\pi^{-}\pi^{0})$ final state, we calculate 
\begin{equation}
	\chi_{\omega\eta}^{2} = \frac{(m_{\pi^{+}\pi^{-}\pi^{0}}^{\omega} - m_{\omega}
	)^2}{\sigma_{\omega}^{2}} + \frac{(m_{\pi^{+}\pi^{-}\pi^{0}}^{\eta} - m_{\eta}
	)^2}{\sigma_{\eta}^{2}}
\end{equation}
where $m_{\pi^{+}\pi^{-}\pi^{0}}^{X}$ $(X=\omega,\eta)$ is the invariant mass of 
the $\pi^{+}\pi^{-}\pi^{0}$ combination for the X candidate and $m_{X}$ is the 
nominal mass of X meson quoted in the PDG.\cite{ParticleDataGroup:2024cfk} We loop over all possible combinations and the 
one with the minimal $\chi_{\omega\eta}^{2}$ value is selected. The polar angle of
the system recoiling against the $\eta$ candidate $\theta_{recoil}$ is also 
required to satisfy $\left|\cos \theta_{recoil}\right| < 0.7$ to minimize the systematic uncertainty. Furthermore, the 
$m_{\pi^{+}\pi^{-}\pi^{0}}^{\omega}$ and $m_{\pi^{+}\pi^{-}\pi^{0}}^{\eta}$ are 
required to be in the ranges [0.65, 0.98] and [0.41, 0.65] $\mathrm{GeV}/c^{2}$,
respectively.

\begin{figure}[t]
\centering
\includegraphics[width=6cm]{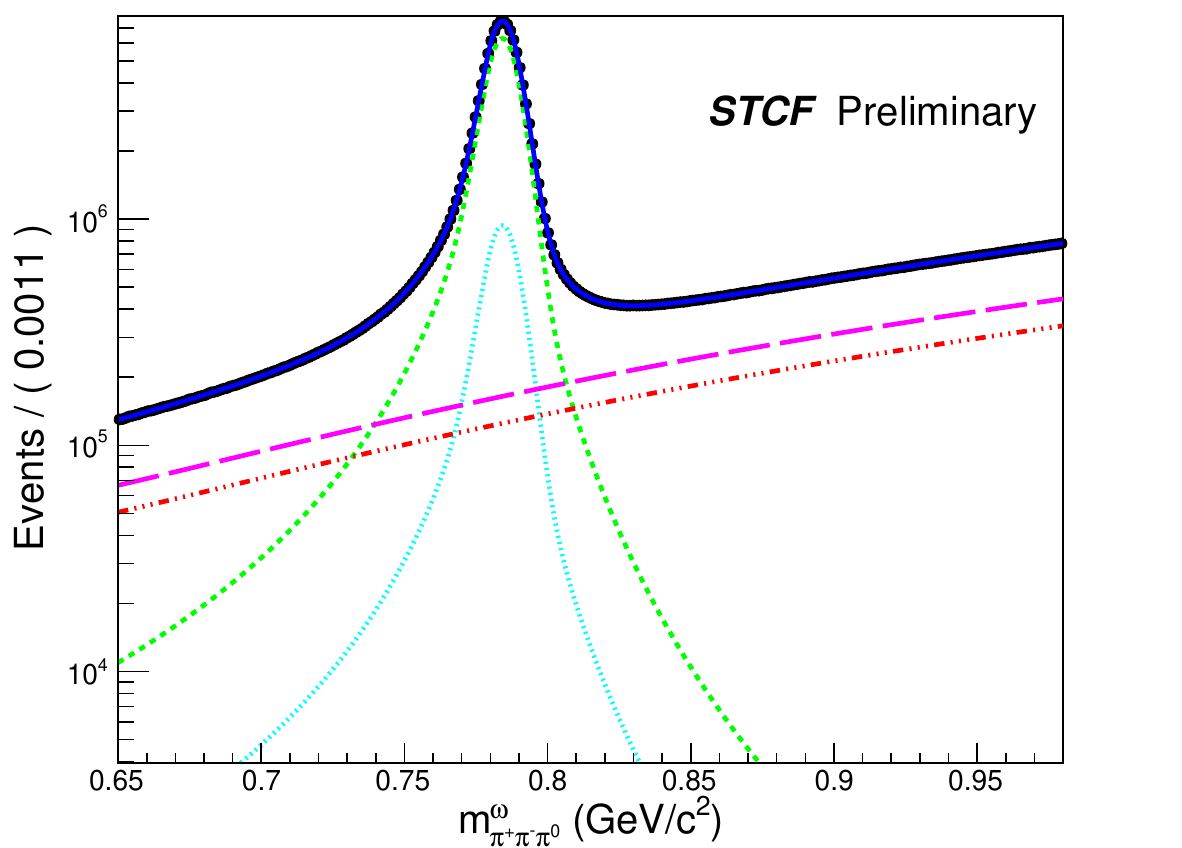}
\includegraphics[width=6cm]{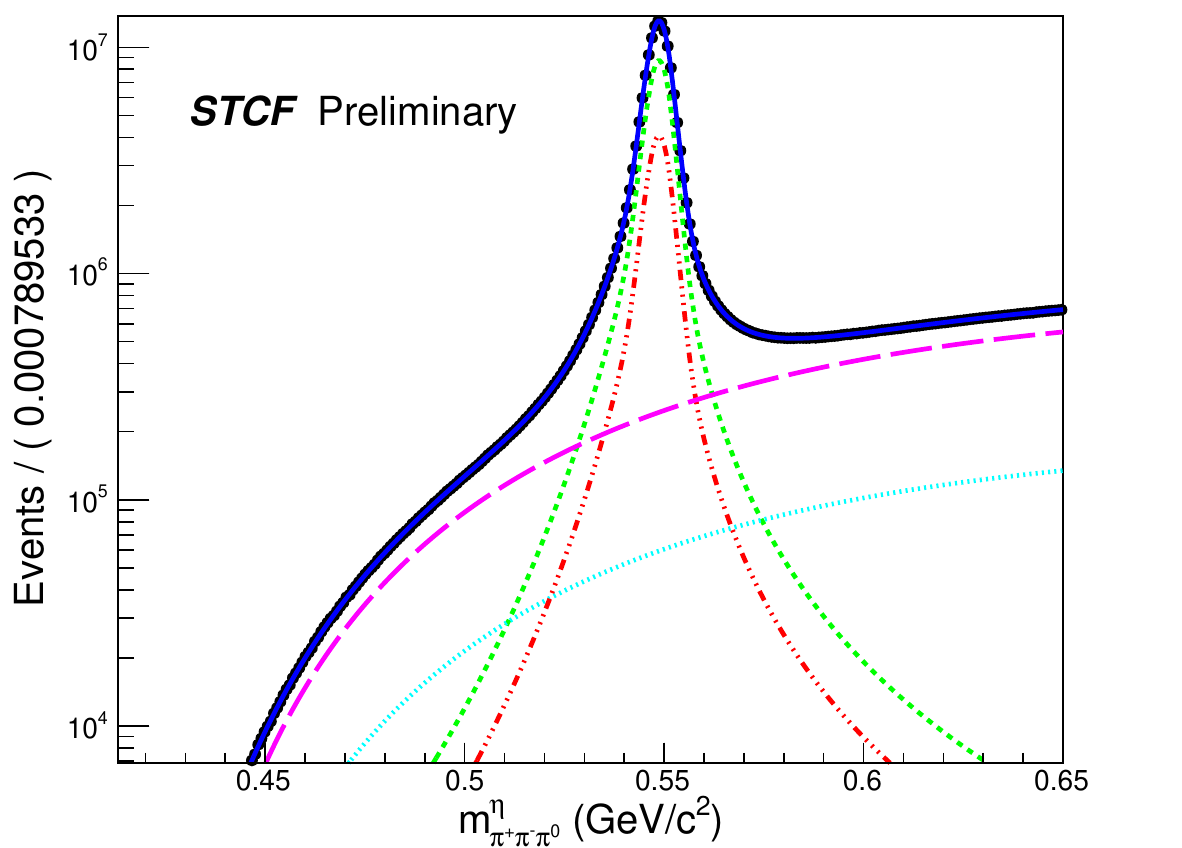}
\caption{Projections of the 2D fit to $m_{\pi^{+}\pi^{-}\pi^{0}}^{\omega}$ and 
$m_{\pi^{+}\pi^{-}\pi^{0}}^{\eta}$ distributions. The projected toy data sample is shown by the 
dots with error bars, the signals by the dashed green curve, BKGI by the 
long-dashed pink curve, BKGII with $\omega$ intermediate state by the dotted cyan 
curve, BKGII with the $\eta$ intermediate state by the dash-dotted red curve, and 
the total fit by the solid blue curve.}
\label{f3}
\end{figure}

The remaining backgrounds are dominated by those with the same final state as the 
signal, but neither $\omega$ nor $\eta$ intermediate states included (named BKGI)
or without one of the two intermediate states (named BKGII). The contributions of 
BKGI and BKGII are determined by performing a 2D ML fit to 
$m_{\pi^{+}\pi^{-}\pi^{0}}^{\omega}$ and $m_{\pi^{+}\pi^{-}\pi^{0}}^{\eta}$. The 
signal and peaking background PDFs are described by the sum of two Crystal Ball 
functions, while the non-peaking background PDF for $\omega$ and $\eta$ candidates 
are described by a second order polynomial and a reversed ARGUS function,\cite{ARGUS:1990hfq} 
respectively.
%  The reversed ARGUS function is defined as 
% \begin{equation}
% 	F(m) = m \cdot (1-(X-m)^{2}/t^{2})^{a} \cdot \mathrm{exp}(-b \cdot 
% 	(1-(X-m)^{2}/t^{2}))
% \end{equation}
% where $X$ is the sum of the lower and upper limits of the fit range, $a$ and $b$ 
% are constant coefficients, and $t$ is the upper limit of the fit range. 
 The signal yield is $N_{sig} = 40801\pm327$ 
and the detection efficiency is 9.1\%. The branching fraction of 
$J/\psi\to\omega\eta$ is measured to be $(1.72\pm0.01)\times10^{-3}$, which is 
consistent with the input value of the inclusive $J/\psi$ MC sample. 
The ML fit yield of the projected toy data is $(106.59 \pm 0.02) \times 10^{6}$.
The projections of the ML fit to the 
$m_{\pi^{+}\pi^{-}\pi^{0}}^{\omega}$ and $m_{\pi^{+}\pi^{-}\pi^{0}}^{\eta}$ 
distributions for toy data are shown in Fig.~\ref{f3}.
% \subsubsection{}
% Typeset subsubheadings in medium face italic and capitalize the
% first letter of the first word only. Section numbers to be in
% roman.

\subsection{The decay mode $J/\psi \to \phi \eta,\phi \to K^{+}K^{-},\eta \to \pi^{+}\pi^{-}\pi^{0}$}
\begin{figure}[t]
\centering
\includegraphics[width=6cm]{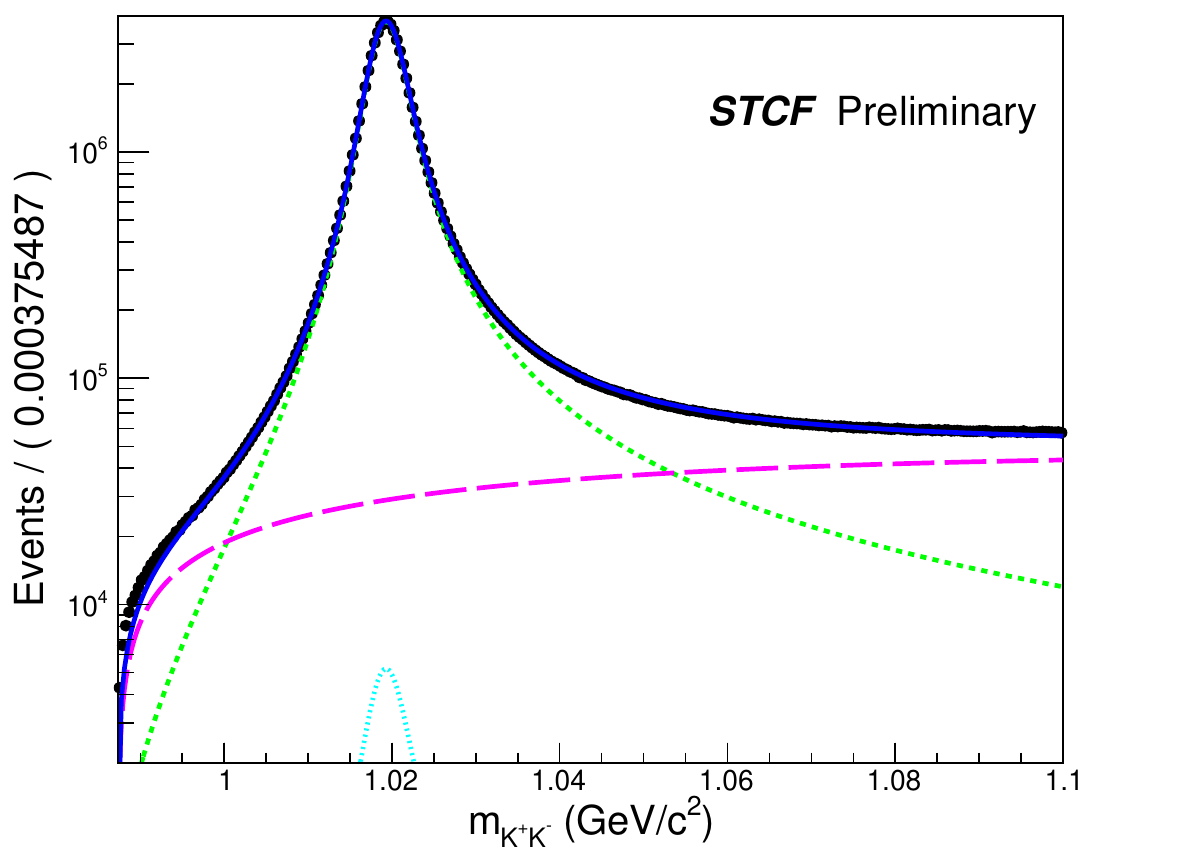}
\includegraphics[width=6cm]{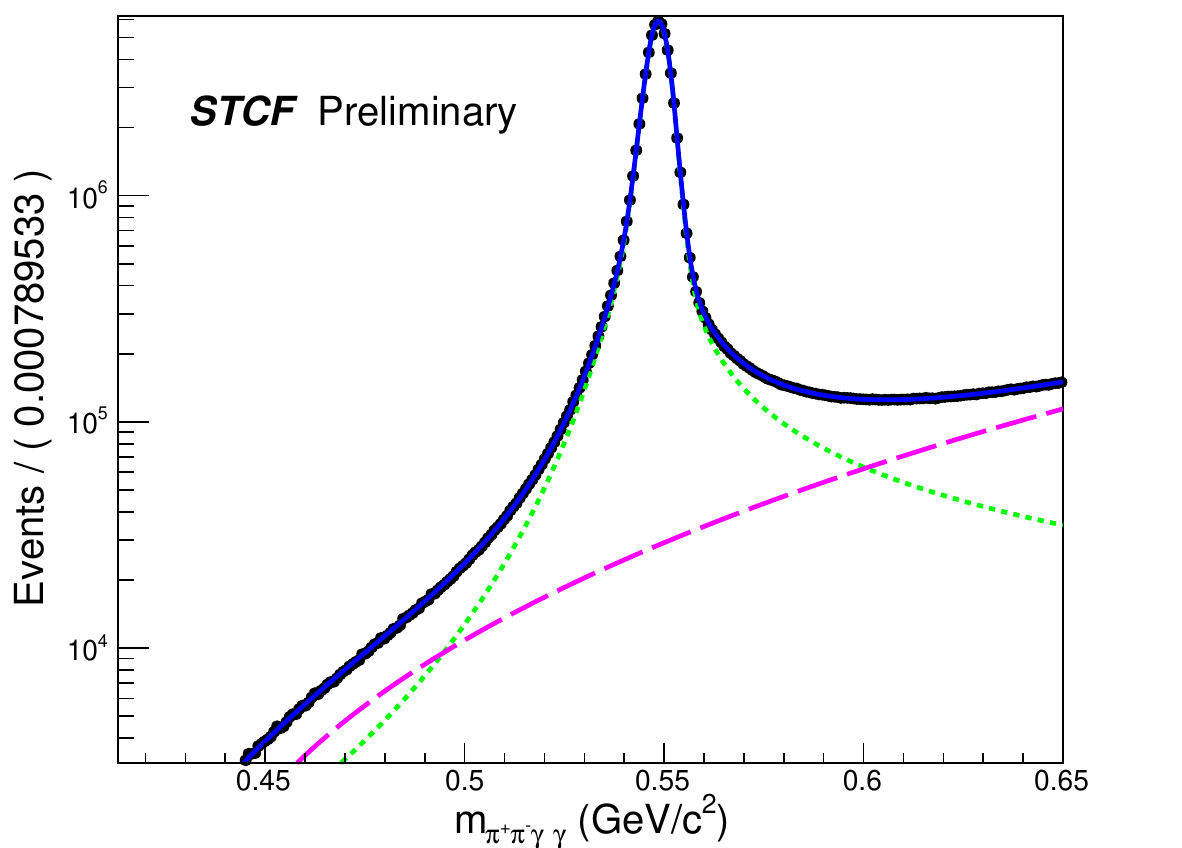}
\caption{Projections of the 2D fit to $m_{K^{+}K^{-}}^{\phi}$ and 
$m_{\pi^{+}\pi^{-}\pi^{0}}^{\eta}$. The projected toy data sample is shown by the 
dots with error bars, the signals by the dashed green curve, BKGI by the 
long-dashed pink curve, BKGII with $\phi$ intermediate state by the dotted cyan 
curve, BKGII with the $\eta$ intermediate state by the dash-dotted red curve, and 
the total fit by the solid blue curve.}
\label{f4}
\end{figure}

For this decay mode, the $\eta$ candidate is reconstructed using the same process as 
that for the invisible decays. The $\phi$ candidate is reconstructed from two 
oppositely charged tracks without a PID requirement. Similar to the visible decay 
mode $\omega \to \pi^{+}\pi^{-}\pi^{0}$, $E^{tot} > 2.95$ GeV and 
$\left|\cos \theta_{recoil}\right| < 0.7$ are also required.
The invariant mass of $K^{+}K^{-}$ is required to be within the range of 
[0.987, 1.10] $\mathrm{GeV}/c^{2}$. The remaining backgrounds are analogous to those 
in the $J/\psi\to\omega\eta$ visible decay.
A 2D ML fit is performed to $m_{K^{+}K^{-}}^{\phi}$ and 
$m_{\pi^{+}\pi^{-}\pi^{0}}^{\eta}$. The signal and peaking background PDF for 
$\phi$ candidate is described by a relativistic Breit-Wigner (BW) function 
convolved with a Gaussian function representing the mass resolution, while the 
signal and peaking background PDFs for $\eta$ candidate are described by the sum of 
two Crystal Ball (CB) functions and the non-peaking background PDFs are described by the 
reversed ARGUS function.\cite{ARGUS:1990hfq} The signal yields
$N_{sig} = 29217\pm196$ and the detection efficiency is 25.85\%. The branching 
fraction of $J/\psi\to\phi\eta$ is measured to be $(7.87\pm0.05)\times10^{-4}$, 
which is consistent with the input value used during the inclusive $J/\psi$ MC sample generation. 
The ML fit yield of the projected toy data is $(75.90 \pm 0.04) \times 10^{6}$.
The $m_{K^{+}K^{-}}^{\phi}$ and $m_{\pi^{+}\pi^{-}\pi^{0}}^{\eta}$ projections for toy data 
are shown in Fig.~\ref{f4}.

\subsection{The decay mode $J/\psi \to \phi \eta(\eta'),\phi \to K^{+}K^{-},\eta(\eta^{\prime}) \rightarrow \gamma\gamma$}

For these decay modes, the selection criteria for the charged tracks are the
same as those for $J/\psi \rightarrow \phi\eta(\eta^{\prime}), \, \eta(\eta^{\prime}) 
\rightarrow invisible$. However, at least two good photons are required. To improve the signal shape, 
a 4-constraint kinematic 
fit is applied to all $J/\psi \rightarrow K^+K^-\gamma\gamma$ hypotheses and the 
combination with the smallest $\chi^2$ is selected. The upper limit on 
$\chi^2$ is set to be 90(40) for the $\eta(\eta^{\prime})$ case. 

\begin{figure}[t]
\centering
\includegraphics[width=6cm]{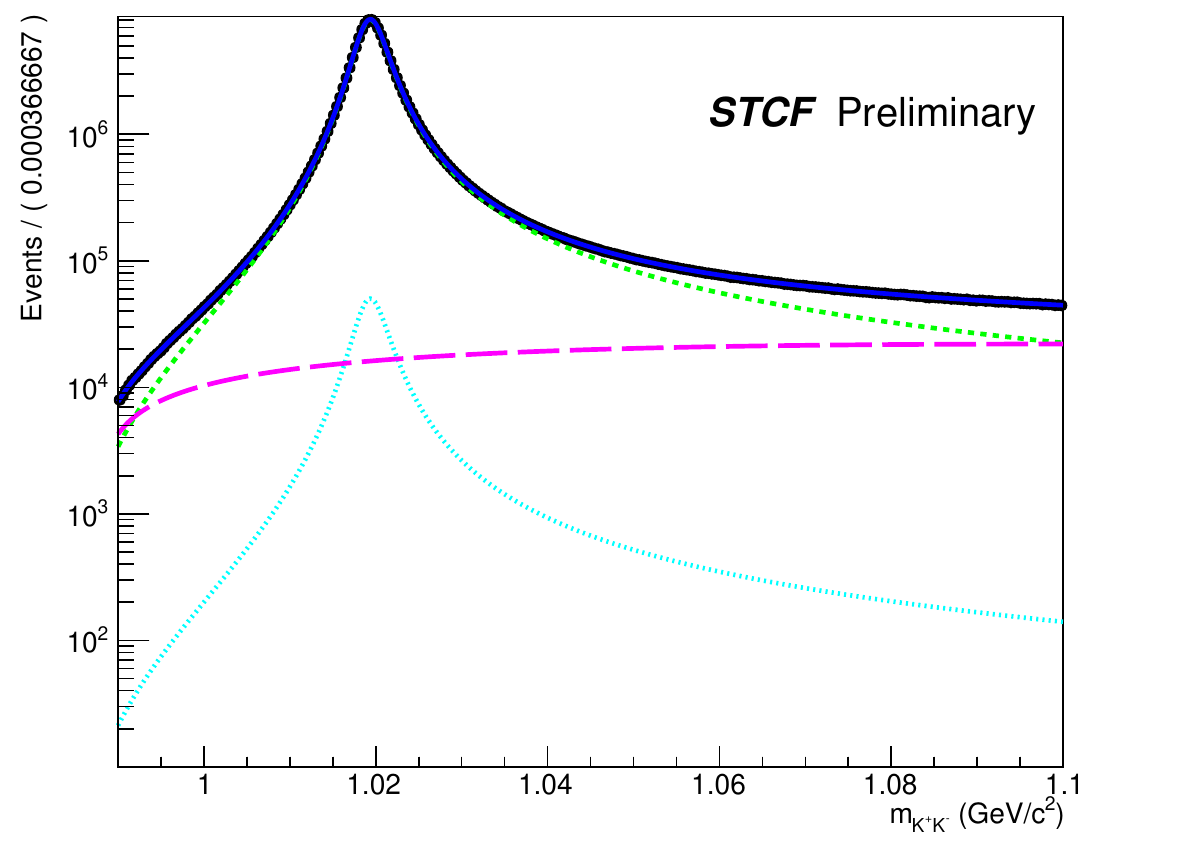}
\includegraphics[width=6cm]{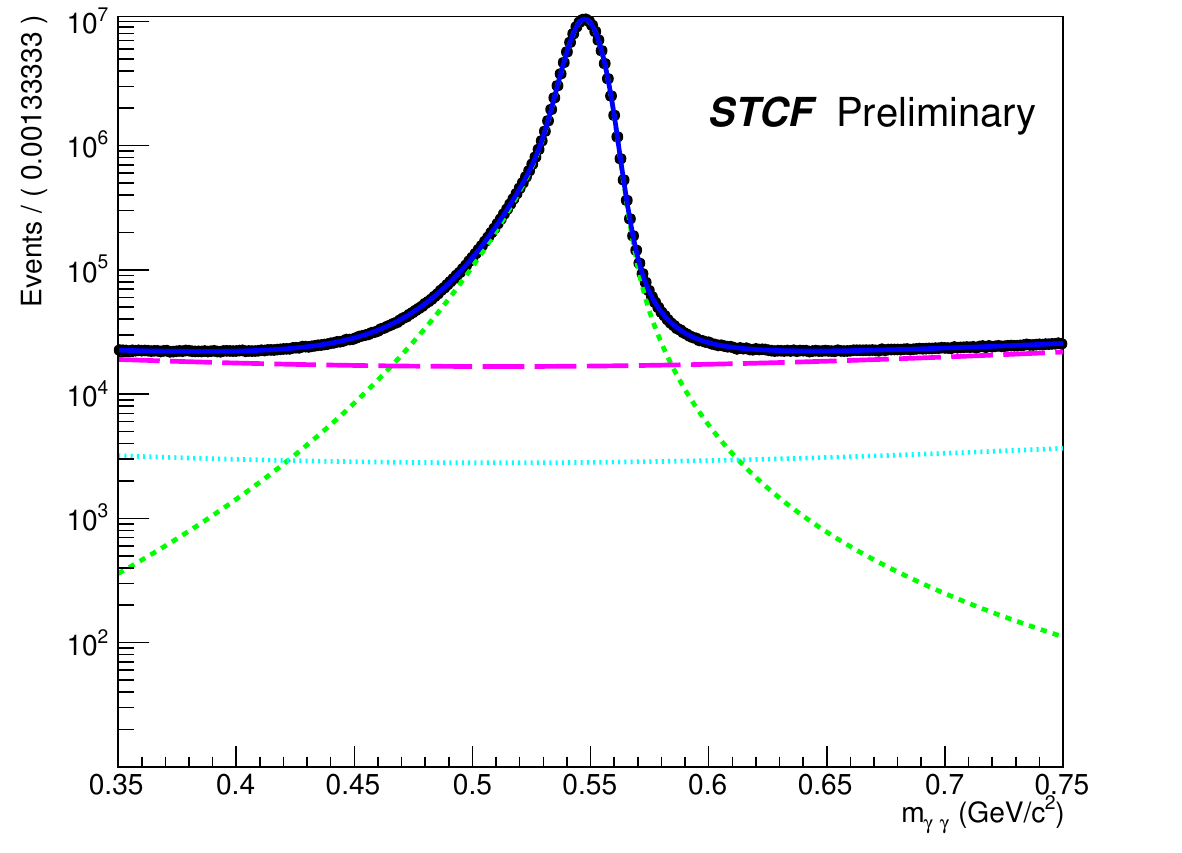}
\caption{Projections of the 2D fit to $m_{K^+K^-}$ and $m_{\gamma\gamma}$. The projected toy data sample is shown by the dots with error bars, 
the signals by the dashed green curve, BKGI by the long-dashed pink curve, BKGII with $\phi$ intermediate state by the dotted cyan curve, BKGII 
with the $\eta$ intermediate state by the dash-dotted red curve, and the total fit by the solid blue curve.}
\label{f7}
\end{figure}

\begin{figure}[t]
\centering
\includegraphics[width=6cm]{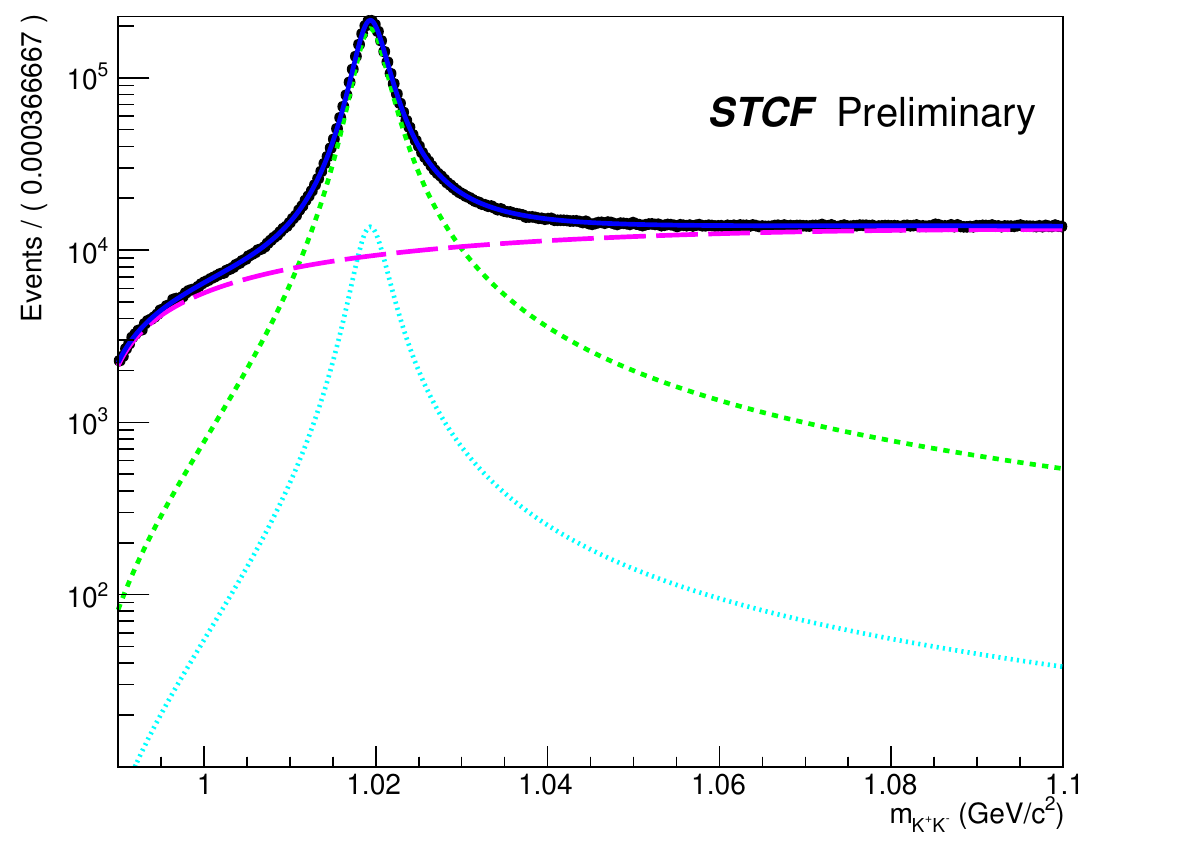}
\includegraphics[width=6cm]{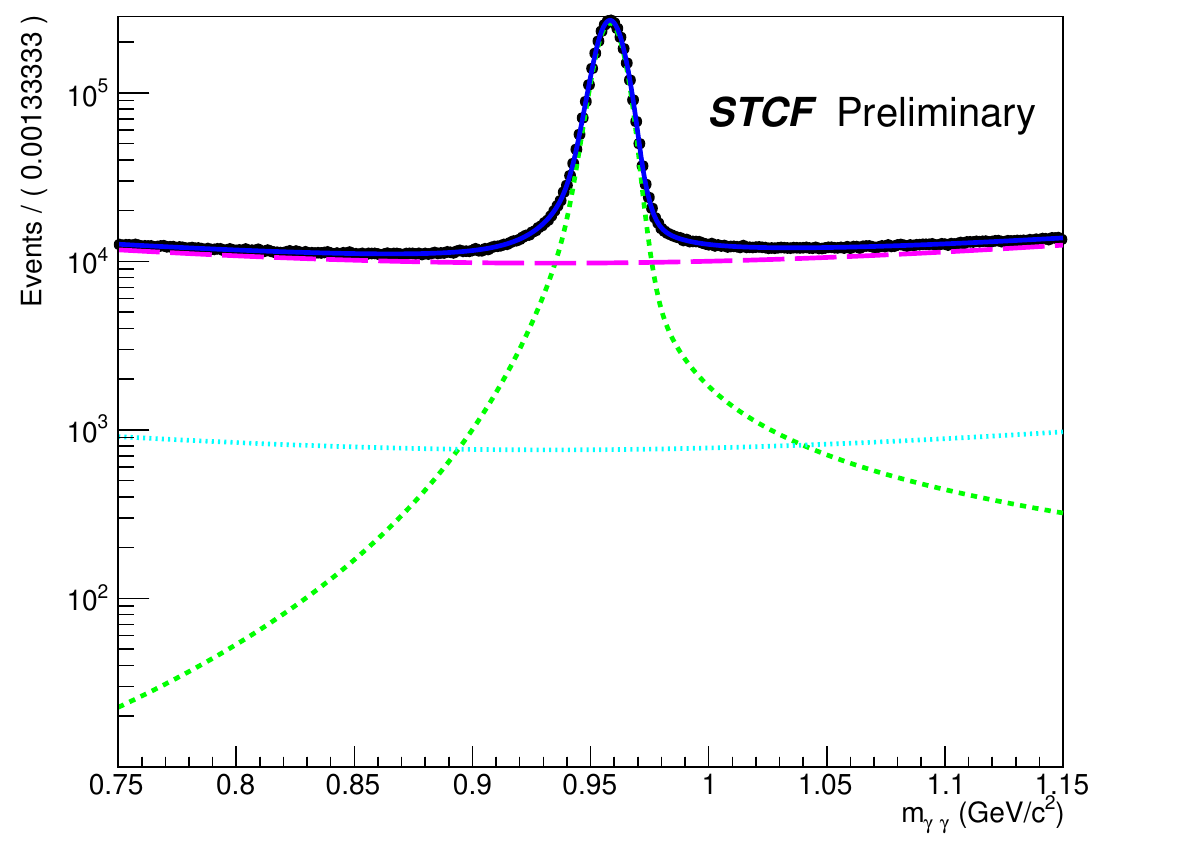}
\caption{Projections of the 2D fit to $m_{K^+K^-}$ and $m_{\gamma\gamma}$. The projected toy data sample is shown by the dots with error bars, 
the signals by the dashed green curve, BKGI by the long-dashed pink curve, BKGII with $\phi$ intermediate state by the dotted cyan curve, BKGII 
with the $\eta^{\prime}$ intermediate state by the dash-dotted red curve, and the total fit by the solid blue curve.}
\label{f8}
\end{figure}

The numbers of $J/\psi \rightarrow \phi\eta(\eta^{\prime}), \, \eta(\eta^{\prime}) 
\rightarrow \gamma\gamma$ events are obtained from an ML fit to the $K^+K^-$ versus 
$\gamma\gamma$ invariant mass distributions. In the ML fits, the invariant mass of $K^+K^-$ is 
required to satisfy $m_{K^+K^-} \in$ [0.99, 1.10] $\mathrm{GeV}/c^{2}$ and the invariant mass of 
$\gamma\gamma$ is required to satisfy $m_{\gamma\gamma} \in$ [0.35, 0.75] $\mathrm{GeV}/c^{2}$ for 
$\eta$ case and $m_{\gamma\gamma} \in$ [0.75, 1.15] $\mathrm{GeV}/c^{2}$ for $\eta^{\prime}$ 
case, respectively. The descriptions of the signal shapes for $\phi$ and $\eta(\eta^{\prime})$ 
are the same as those for the visible decay of $\phi$ meson. The non-peaking background PDFs for $\phi$ 
and $\eta(\eta')$ candidates are described by a reversed ARGUS function\cite{ARGUS:1990hfq} and a second 
order polynomial, respectively. The signal yields are $N_{sig}^{\eta \rightarrow \gamma\gamma} = 
55490 \pm 241$ and $N_{sig}^{\eta^{\prime} \rightarrow \gamma\gamma} = 1335 \pm 48$. 
By considering the corresponding signal detection efficiencies $\epsilon^{\eta \rightarrow 
\gamma\gamma} = 29.56\%$ and $\epsilon^{\eta^{\prime} \rightarrow \gamma\gamma} = 21.92\%$, 
the calculated branching fractions of $J/\psi \rightarrow \phi\eta$ and 
$J/\psi \rightarrow \phi\eta^{\prime}$ are $(7.51 \pm 0.03) \times 10^{-4}$ and 
$(4.39 \pm 0.16) \times 10^{-4}$, which are consistent with the input with the input value of the 
inclusive $J/\psi$ MC sample. After generating 
projected toy data, the signal events obtained are $N^{\eta \rightarrow \gamma\gamma} = (145.16 \pm 0.01) \times 10^{6}$ and 
$N^{\eta^{\prime} \rightarrow \gamma\gamma} = (34.85 \pm 0.02)\times 10^{5}$ are obtained. 
The fit results for $m_{K^+K^-}$ and $m_{\gamma\gamma}$ on toy data 
are shown in Fig.~\ref{f7} and Fig.~\ref{f8} for the $\eta$ and
$\eta^{\prime}$ cases, respectively.

% `Substituting the results into the equation(1) will yield, 

% \centering
% \begin{tabular}{c|c}
% 	\hline
% 	\hline
% 	ratio & at 90\% C.L. \\ 
% 	\hline
% 	$\frac{\mathcal{B}(\eta \to invisible)}{\mathcal{B}(\eta \to \gamma\gamma)}$ & $4.69 \times 10^{-7}$ \\ 
% 	\hline
% 	$\frac{\mathcal{B}(\eta^{\prime} \to invisible)}{\mathcal{B}(\eta^{\prime} \to \gamma\gamma)}$ & $1.76 \times 10^{-5}$ \\ 
% 	\hline
% 	\hline
% \end{tabular}'

\section{Results and outlook}
With the generated projected toy data, the upper limits on the signal events of the invisible decays of $\omega/\phi/\eta/\eta'$ 
mesons at the 90\% confidence level (C.L.) are computed using the Bayesian approach.\cite{ParticleDataGroup:2024cfk} 
The branching ratios of $\frac{\mathcal{B}(\omega \to invisible)}{\mathcal{B}(\omega \to \pi^{+}\pi^{-}\pi^{0})}$, 
$\frac{\mathcal{B}(\phi \to invisible)}{\mathcal{B}(\phi \to K^{+}K^{-})}$, 
$\frac{\mathcal{B}(\eta \to invisible)}{\mathcal{B}(\eta \to \gamma\gamma)}$ and 
$\frac{\mathcal{B}(\eta' \to invisible)}{\mathcal{B}(\eta' \to \gamma\gamma)}$ are calculated using 
Eq.~\eqref{e1}. By using the branching fractions of $\omega \to \pi^{+}\pi^{-}\pi^{0}$,
$\phi \to K^{+}K^{-}$, $\eta \to \gamma\gamma$ and $\eta' \to \gamma\gamma$ quoted in the PDG,\cite{ParticleDataGroup:2024cfk} the upper limits 
on the invisible decay branching fractions at 90\% C.L. are calculated. The results and theoretical predictions\cite{mcelrath2007lighthiggsesdarkmatter} are summarized in Table~\ref{ta1}.

By following traditional methods the upper limits on the invisible decay branching fractions are expected to reach 
$\sim 10^{-7}$ according to our study, while the theoretical predictions of the invisible decay branching fractions 
is $\sim 10^{-8}$ for $\omega$ and $\phi$ mesons.\cite{mcelrath2007lighthiggsesdarkmatter} We propose to use machine learning techniques such as Deep Learning,\cite{vaswani2023attentionneed,qu2024particletransformerjettagging}
which may help reduce backgrounds and enhance sensitivity. Thus the final results may bring more confidence in 
searching for invisible decays and studying the nature of dark matter at STCF.

\begin{table}[t]
	\tbl{The 90\% C.L. upper limits on the branching ratios and invisible decay branching fractions of $\omega$ 
	($\phi$) and $\eta$ ($\eta'$) mesons, compared with the corresponding theoretical predictions for the branching fractions.\label{ta1}}
	{\begin{tabular}{c|c|c|c|c}
		\hline
		\hline
		Observable & $\omega$ & $\phi$ & $\eta$ & $\eta'$ \\
		\hline
		Reference visible decay & $\omega \to \pi^{+}\pi^{-}\pi^{0}$ & $\phi \to K^{+}K^{-}$ & $\eta \to \gamma\gamma$ & $\eta' \to \gamma\gamma$ \\
		\hline
		$\frac{\mathcal{B}(X \to invisible)}{\mathcal{B}(X \to visible)}$ & $4.2\times 10^{-7}$ & $1.8\times 10^{-6}$ & $4.7\times 10^{-7}$ & $1.8\times 10^{-5}$ \\
		\hline 
		$\mathcal{B}(X \to invisible)$ & $3.7\times 10^{-7}$ & $8.9\times 10^{-7}$ & $1.8\times 10^{-7}$ & $4.1\times 10^{-7}$ \\
		\hline
		Theoretical prediction & $7.2 \times 10^{-8}$ & $1.9 \times 10^{-8}$ & $3.4 \times 10^{-5}$ & $3.7 \times 10^{-7}$ \\
		\hline
		\hline
	\end{tabular}}
\end{table}

\section*{Acknowledgments}
% This section should come before the References. Dedications and funding
% information may also be included here.
We thank the supercomputing center of Lanzhou University for their strong support. We thank Yu Zhang and Mingyi Liu, convenors of the STCF basic
symmetries physics simulation group. We are also grateful to the STCF hardware and software teams for their contributions to the detector simulation
and offline software framework used in this work. This work is supported by the National Key R\&D Program of China under Contracts No. 2022YFA1602200 and No. 2023YFA1607200; the National
Natural Science Foundation of China (NSFC) under Contracts No. 12341501, No. 12341503, No. 12341504; the international partnership program 
of the Chinese Academy of Sciences Grant No. 211134KYSB20200057 and the Seed Funding of Jilin University. 
We thank the Hefei Comprehensive National Science Center for their strong support on the STCF key technology research project.

\section*{ORCID}
% You are encouraged to include in your user information the ORCID (\url{https://orcid.org/}) or register for one if you don't have it. This ID will help to identify you in the researcher community and make it easier to keep track of all your publications.
% Please provide a valid ORCID here, e.g.,

\noindent Zhibo Tao - \url{https://orcid.org/0009-0006-0753-0479}

% \noindent Rajesh Babu - \url{https://orcid.org/0009-0006-0415-6880}

\appendix

\bibliographystyle{ws-ijmpa}
\bibliography{ref}
\end{document}